\documentclass{article}

\usepackage{PRIMEarxiv}

\usepackage[utf8]{inputenc} 
\usepackage[T1]{fontenc}    
\usepackage{hyperref}       
\usepackage{url}            
\usepackage{booktabs}       
\usepackage{amsfonts}       
\usepackage{nicefrac}       
\usepackage{microtype}      

\usepackage{fancyhdr}       
\usepackage{graphicx}       
\graphicspath{{Figures/}}   

\usepackage[compress]{cite}
\usepackage{amsmath,amssymb,amsfonts}
\usepackage{stfloats}
\usepackage[caption=false]{subfig}
\usepackage{array}

\title{
    Collaboration in Multi-Robot Systems:
    \break Taxonomy and Survey of Frameworks For Collaboration
}

\author{
  Riwa Karam\(^1\), Alexander A. Nguyen\(^1\), Ruoyu Lin\(^1\), David R. Martin\(^1\), \\ \textbf{Diana Morales}\(^1\), \textbf{Brooks A. Butler}\(^2\), \textbf{Magnus Egerstedt}\(^3\) \\
  \(^1\)University of California, Irvine, Irvine, CA, USA \\
  \(^2\)Oklahoma State University, Stillwater, OK, USA \\
  \(^3\)University of North Carolina at Chapel Hill, Chapel Hill, NC, USA \\
  \texttt{Email:\{rwkaram, alexaan2, rlin10, davidrm3, dlmoral2\}@uci.edu}, \\ \texttt{brooks.butler@okstate.edu, magnus@unc.edu}
}

\newtheorem{definition}{Definition}

\begin{document}
\maketitle

\begin{abstract}
    Collaboration is a central theme in multi-robot systems as tasks and demands increasingly require capabilities that go beyond what any one individual robot possesses. Yet, despite extensive work on cooperative control and coordinated behaviors, the terminology surrounding collective multi-robot interaction remains inconsistent across research communities. In particular, cooperation, coordination, and collaboration are often treated interchangeably, without clearly articulating the differences among them. To address this gap, we propose definitions that distinguish and relate cooperation, coordination, and collaboration in multi-robot systems, highlighting the support of new capabilities in collaborative behaviors, and illustrate these concepts through representative examples. Building on this taxonomy, different frameworks for collaboration are reviewed, and technical challenges and promising future research directions are identified for collaborative multi-robot systems.
\end{abstract}


\section{Introduction}\label{sec:introduction}

Many of today’s engineering challenges, where robots are envisioned to be utilized, such as disaster response \cite{Murphy17:Disaster}, environmental monitoring \cite{Dunbabin12:Robots}, large-scale logistics \cite{Karabegovic15:The_Application}, and space exploration \cite{Gao17:Review}, are well-suited to multi-robot deployments, particularly as tasks grow beyond the capabilities of any single robot. These applications call out the need for multi-robot \textit{collaboration}, where robots share goals and combine their different functionalities (such as, sensing and actuation modalities, or other physical attributes) to achieve results that are difficult or impossible to achieve in isolation. As a result, collaboration has become a central theme in multi-robot research \cite{Stroupe05:Behavior, Miller22:Stronger, Zhou2022, Nguyen23:Mutualistic, RoCo2024, Butler25:Collaborative}, prompting the need to understand how such joint behaviors should be described and classified.

Despite the prevalence of multi-robot systems, the terminology used to describe collective behaviors remains inconsistent across research communities. One contributing factor is that related concepts, such as "cooperation" or "coordination", are by themselves broad and variably interpreted, making it difficult to delineate finer-grained categories of cooperative behavior in a way that is clearly defined and technically precise. As a result, "cooperation", "coordination", and "collaboration" are frequently invoked in robotics, and in multi-agent systems broadly \cite{farinelli2004multirobot, yan2013survey, Rizk20:Cooperative}, but not clearly defined or distinguished. These terms often appear interchangeably, even though they may reflect distinct assumptions regarding inter-robot dependencies,
information exchanges, and the nature of joint actions among robots. For example, the standard approaches to consensus \cite{Mesbahi10:Graph, amirkhani2022consensus}, formation \cite{Tabuada01:Feasible, Chen05:Formation}, and coverage \cite{Cortes04:Coverage, cortes2002coverage} typically emphasize the notion of coordination, which ensures that individual actions remain compatible with a shared objective, without necessarily enabling capabilities that surpass those of individual robots. Thus, as multi-robot systems become more heterogeneous and tasks increasingly coupled and complex, the lack of shared definitions complicates comparisons across methods and may obscure how cooperative approaches might transfer across problem settings and team compositions.

While several surveys have reviewed "cooperative" or "coordinated" multi-robot systems \cite{Cortes17:Coordinated, Ismail18:A_Survey, Chen25:A_Survey}, and others have examined resilience \cite{Zhang17:Resilient, prorok2021beyond} and task allocation \cite{Gerkey04:A_Formal, korsah13:Comprehensive}, or domain-specific deployments (such as, object transportation \cite{Tuci18:Cooperative, An23:Multi_Robot}), there remains limited treatment of "collaboration" as a distinct and carefully defined concept. Existing works typically focus on particular subsets of the problem (for example, industrial collaboration architectures \cite{menebroker2025multi}, cooperative control methods \cite{Rizk20:Cooperative}, or multi-agent learning techniques \cite{Orr23:Multi_Agent}) without addressing how the different approaches relate or what conditions are necessary for the "collaborative" capabilities to be supported.

A common vocabulary may also support practical comparison across studies. Researchers choosing a method may need to understand whether it helps robots divide work that they can perform individually or enables them to accomplish a task that requires their combined abilities. The assumptions each method makes about the robots and the information available to them may also influence this choice. Making these distinctions explicit could offer a consistent basis for comparing approaches developed for different applications, identifying which parts of a solution might be reused, and considering what additional mechanisms a new task may require. In this way, the taxonomy may complement the detailed problem formulations in individual papers by helping readers examine their implications for system design across the literature.

The objective of this article is to address this lack of clear distinctions among cooperation, coordination, and collaboration, by providing a structured examination of collaboration in multi-robot systems. This article provides the following main contributions:
\begin{itemize}
    \item Proposed definitions of cooperation, coordination, and collaboration and clarification of their relationships;
    \item Review of different organizational architectures (centralized, decentralized, hierarchical) supporting collaborative behavior; and
    \item Review of collaborative frameworks drawn from ecology, human-swarm interaction, game theory, and learning, together with additional approaches based on task planning and system integration.
\end{itemize}
By synthesizing insights from multiple perspectives within the literature on collaborative multi-robot systems, this survey aims to establish a conceptual foundation and to support the development of new methodologies that explicitly leverage collaborative capabilities in multi-robot systems.

The outline of this article is as follows. The next section presents the definitions adopted in this survey and illustrates them through representative examples. The "Background on Cooperation and Coordination" Section reviews the background literature. The "Multi-Robot Collaboration" Section examines organizational structures, four selected framework inspirations and tools, and additional collaborative approaches from task planning and system integration. The "Comparisons" Section summarizes their organizational and methodological features. The "Challenges and Future Research Directions" Section discusses open questions and opportunities for collaborative multi-robot systems.

\section{Taxonomy on Cooperation, Coordination, and Collaboration} \label{sec:taxonomy}

\begin{figure}[!t]
    \centering
    \includegraphics[width=0.5\columnwidth]{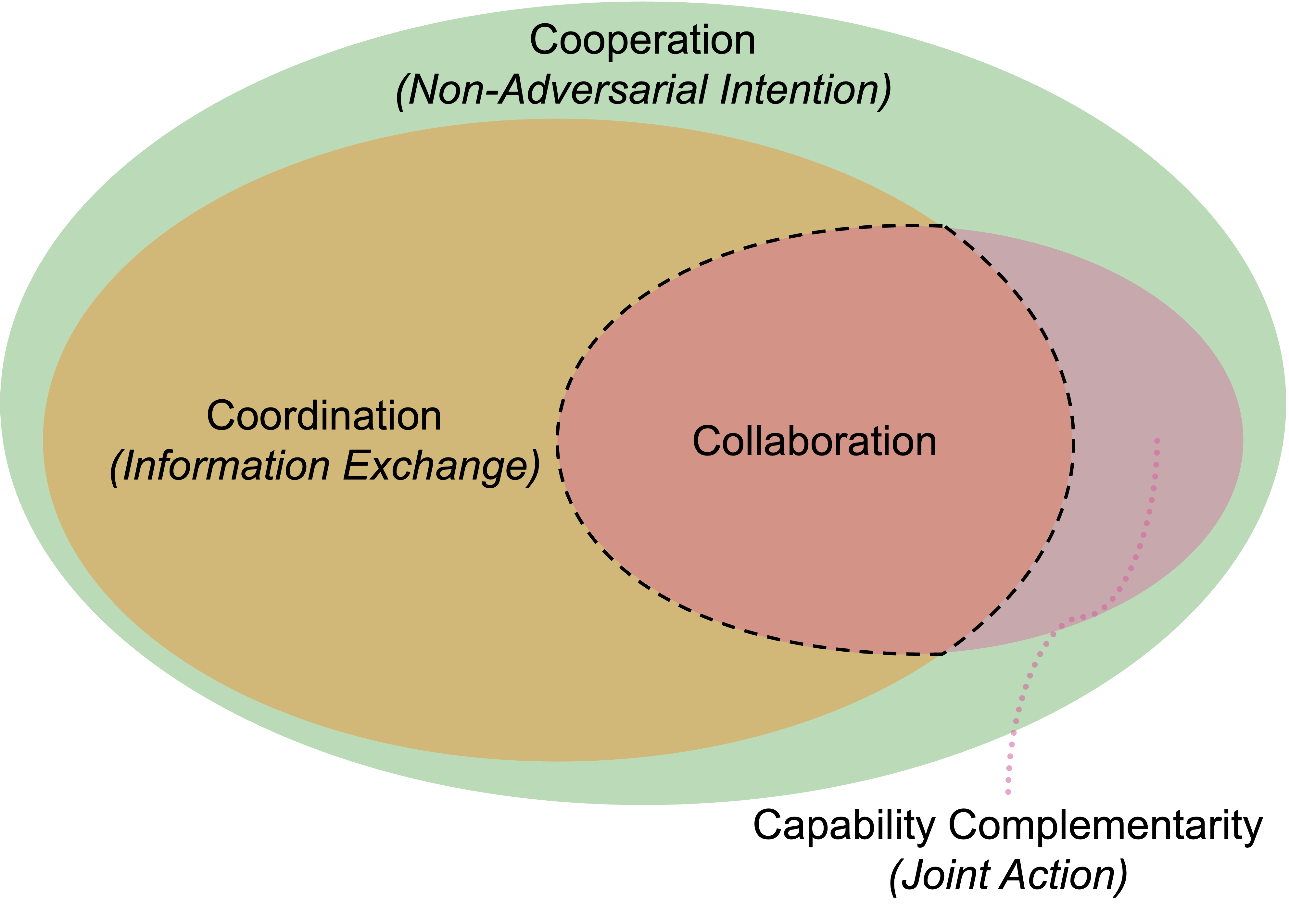}
    \caption{Venn diagram illustrating the relationship among cooperation, coordination, and collaboration in multi-robot systems. Both coordination and capability complementarity are subsets of cooperation. Capability complementarity includes both its overlap with coordination, which constitutes collaboration, and the portion outside coordination, which represents uncoordinated capability complementarity. Collaboration is therefore part of capability complementarity, not a separate set.}
    \label{fig:venn_diagram}
\end{figure}

Given the potential for ambiguity in the terminology used to describe collective behaviors in multi-robot systems, we introduce a set of definitions in this section that distinguish cooperation, coordination, and collaboration, as well as the relationship between these terms. These concepts can often appear together in the literature, but may represent fundamentally different types of inter-robot relationships, as well as the extent to which robots rely on one another to complete a task. The taxonomy introduced below, visualized in Figure~\ref{fig:venn_diagram}, will serve as a basis for this article.

The definitions below are the terminology adopted for this survey, synthesized from the task objectives and interaction mechanisms discussed in the existing literature. Surveys of multi-robot coordination and cooperative heterogeneous systems organize related work around communication, task structure, and robot capabilities \cite{yan2013survey, Rizk20:Cooperative}. The consensus, formation, and coverage formulations reviewed in the next section illustrate shared objectives and information-dependent actions \cite{olfati2004consensus, olfati2002graph, Cortes04:Coverage}. The transport and perception frameworks discussed later illustrate the additional requirement that robots combine capabilities to achieve a task outcome unavailable to any individual robot under the same task assumptions \cite{zhang2020decentralized, Zhou2022}. Our taxonomy makes these distinctions explicit through cooperation, coordination, and capability complementarity. It classifies the mechanisms and task requirements described in a work, rather than the choice of ``cooperative,'' ``coordinated,'' or ``collaborative'' in its title; the definitions are not intended to replace the terminology used by the original authors.

This survey focuses on cooperation, coordination, and collaboration in multi-robot systems. The choice is motivated by the aim of understanding how robots combine their capabilities to accomplish tasks together: under the definitions adopted here, collaboration builds on cooperation and coordination. The scope therefore centers on non-adversarial relationships among robots. Competitive and adversarial interactions belong to a broader body of multi-agent research \cite{yi2022survey, zhang2021multi, lowe2017multi}; a taxonomy of those interactions, including antagonistic behavior or deliberate sabotage, is outside the scope of this survey. Other perspectives, such as resilience and self-organization, may intersect with the relationships considered here, but are not comprehensively classified in this article.

Within this scope, the taxonomy describes relationships among robot behaviors rather than an exhaustive partition of multi-agent application domains. As Figure~\ref{fig:venn_diagram} illustrates, its categories are nested and intersecting: a collaborative behavior is also cooperative and coordinated. Task constraints, robot heterogeneity, organizational structure, and reliability requirements provide additional attributes of a problem without each requiring a new interaction category. The criteria are applied to a specified task and its requirements; different activities or phases within one application can therefore exhibit different relationships. This scope may support comparison across applications while retaining the problem-specific detail of their formulations.

Cooperation represents one of the more fundamental forms of collective behavior, requiring only that robots share an overarching objective and that they do not act in ways that deliberately impede one another. We present the definition of cooperation in Definition~\ref{def:cooperation}; note that, already, we are imposing a notion of "intention" for agents in robotic systems, which in and of itself highlights the challenge in defining behavioral terms for engineered systems where intent and agency in autonomous systems can have broad interpretations. However, for the purposes of this article, we consider the class of cooperative multi-robot systems to include all scenarios where the \textit{intention} is that robotic agents either work towards a shared objective, or at the very least do not explicitly make efforts to the detriment of other agents.

\begin{definition} \label{def:cooperation}
    \textbf{Cooperation} between two or more robots is the \textbf{non-adversarial intention} to contribute towards achieving a shared goal or a set of related tasks.
\end{definition}

The definition of cooperation does not impose any constraints on the information exchange, task execution, or joint effort between robots. It imposes minimal assumptions on robot interaction rules, serving as a foundation for richer forms of collective behavior introduced later in this section. In the case of system behavior that is only cooperative, but not necessarily coordinated or collaborative, robots may operate independently, without coupling their decisions or actions, as long as their behaviors are aligned with their shared goal or related tasks. Cooperative robots may also use information about one another to guide their actions. This information may be conveyed through explicit messages, including wireless communication, or obtained by locally sensing other robots' states and actions. Both pathways are included in the notion of information exchange used in this taxonomy. When such information couples robot decisions, it introduces the additional structure associated with coordination (Definition~\ref{def:coordination}).

\begin{figure*}[!t]
    \centering
    \subfloat[Cooperation]{
        \includegraphics[width=0.485\textwidth]{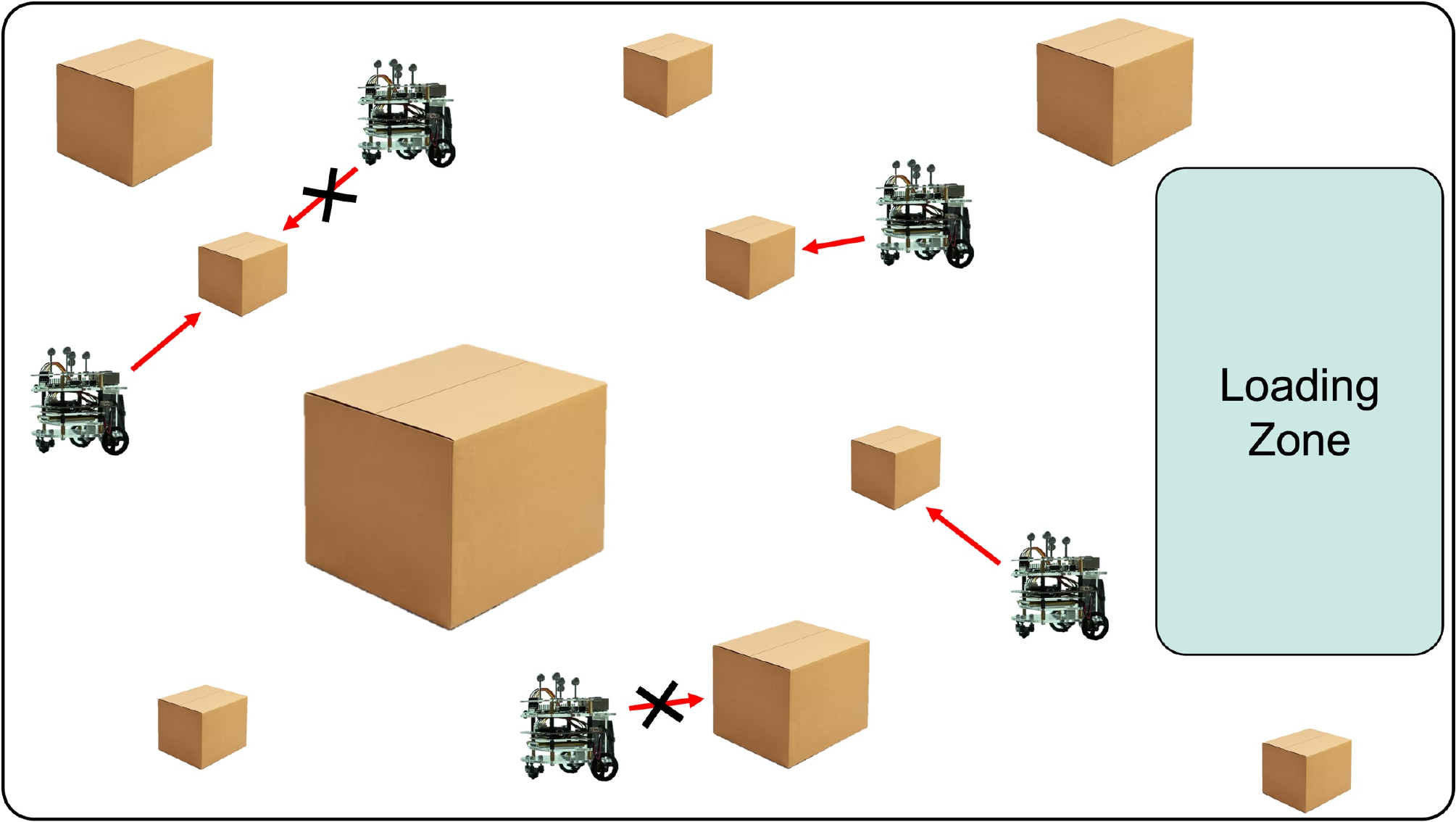}
        \label{fig:coop}
    }
    \hfill
    \subfloat[Coordination]{
        \includegraphics[width=0.485\textwidth]{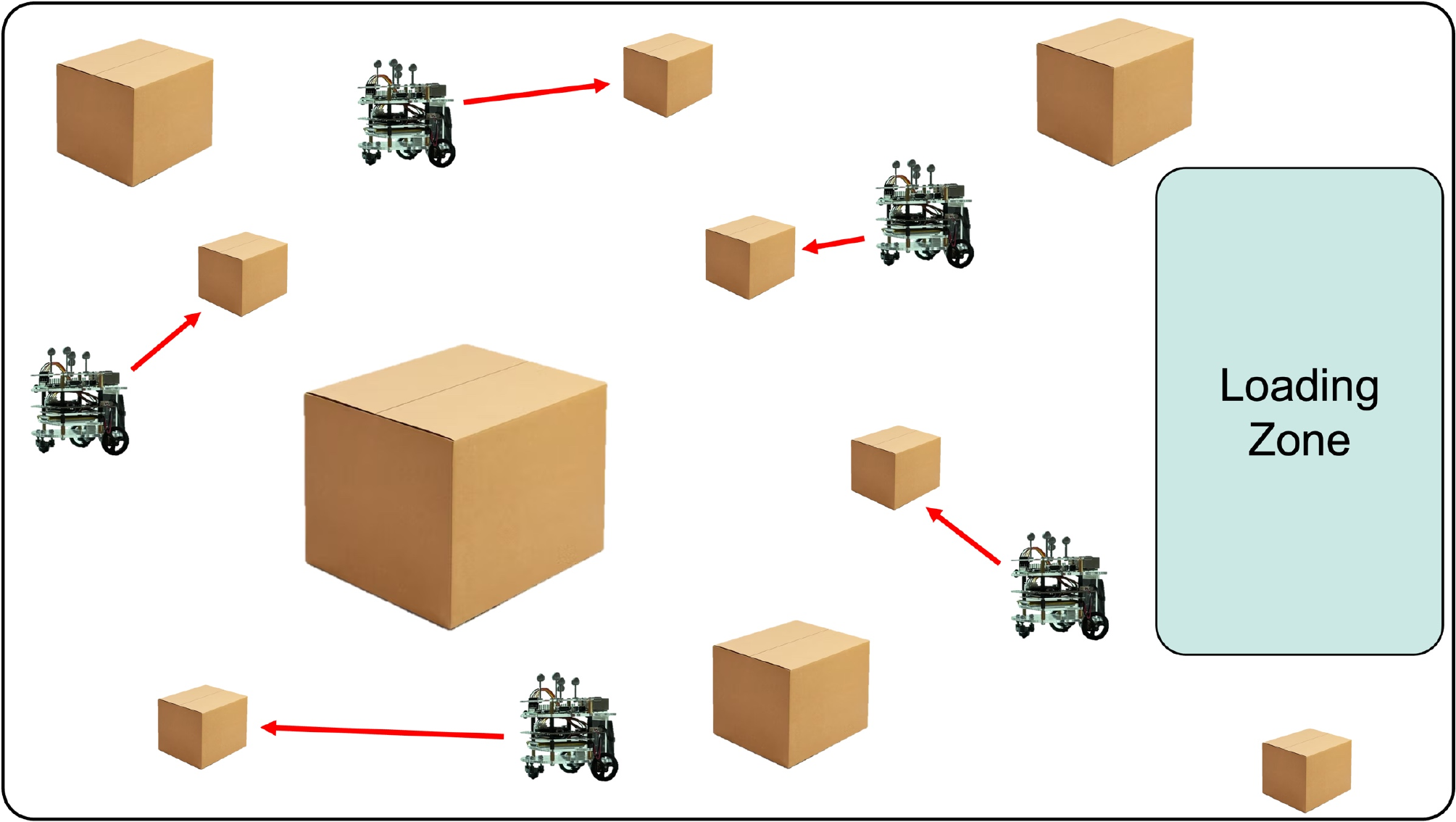}
        \label{fig:coord}
    }
    \hfill
    \subfloat[Capability Complementarity]{
        \includegraphics[width=0.485\textwidth]{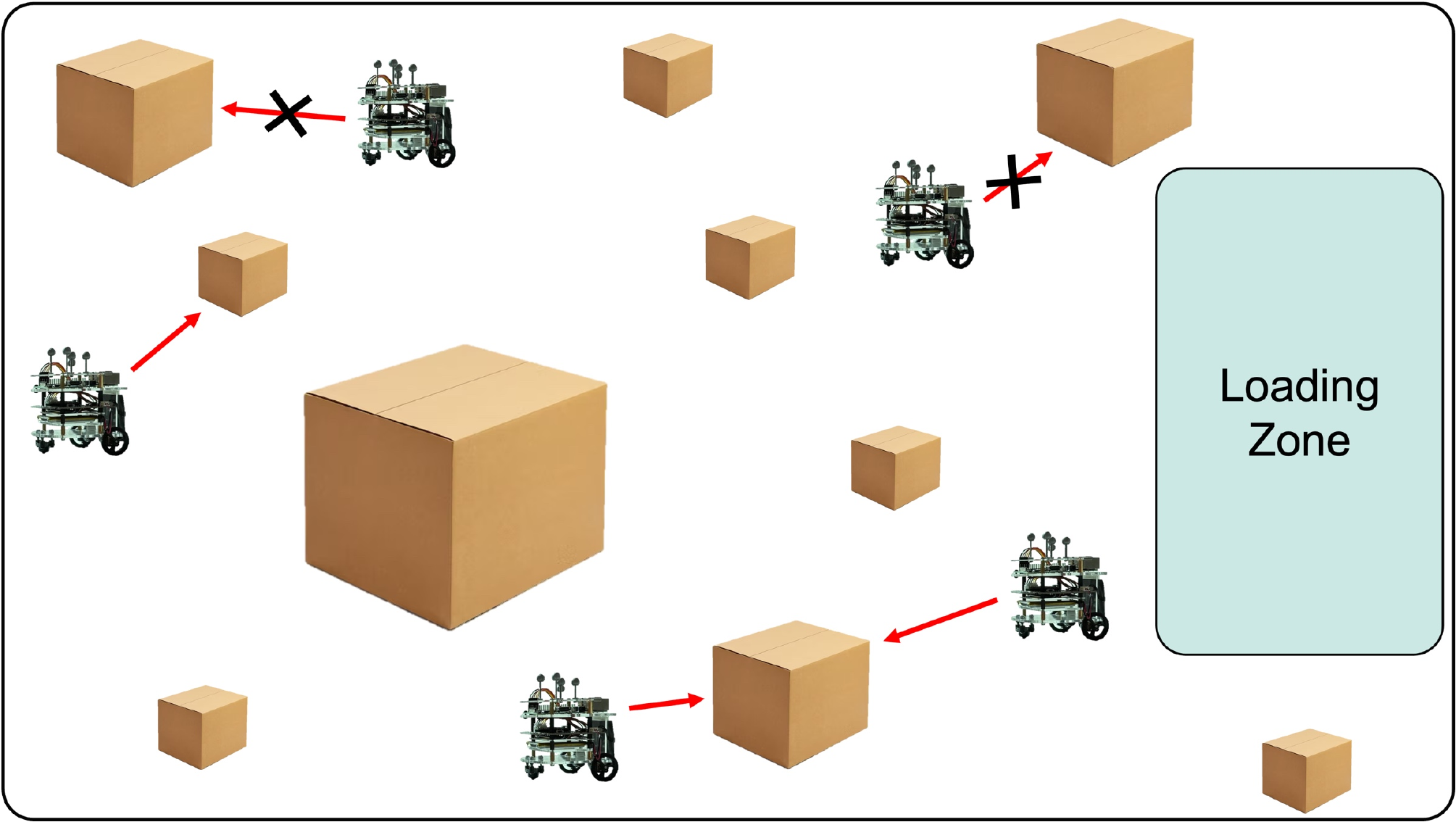}
        \label{fig:capab_comp}
    }
    \hfill
    \subfloat[Collaboration]{
        \includegraphics[width=0.485\textwidth]{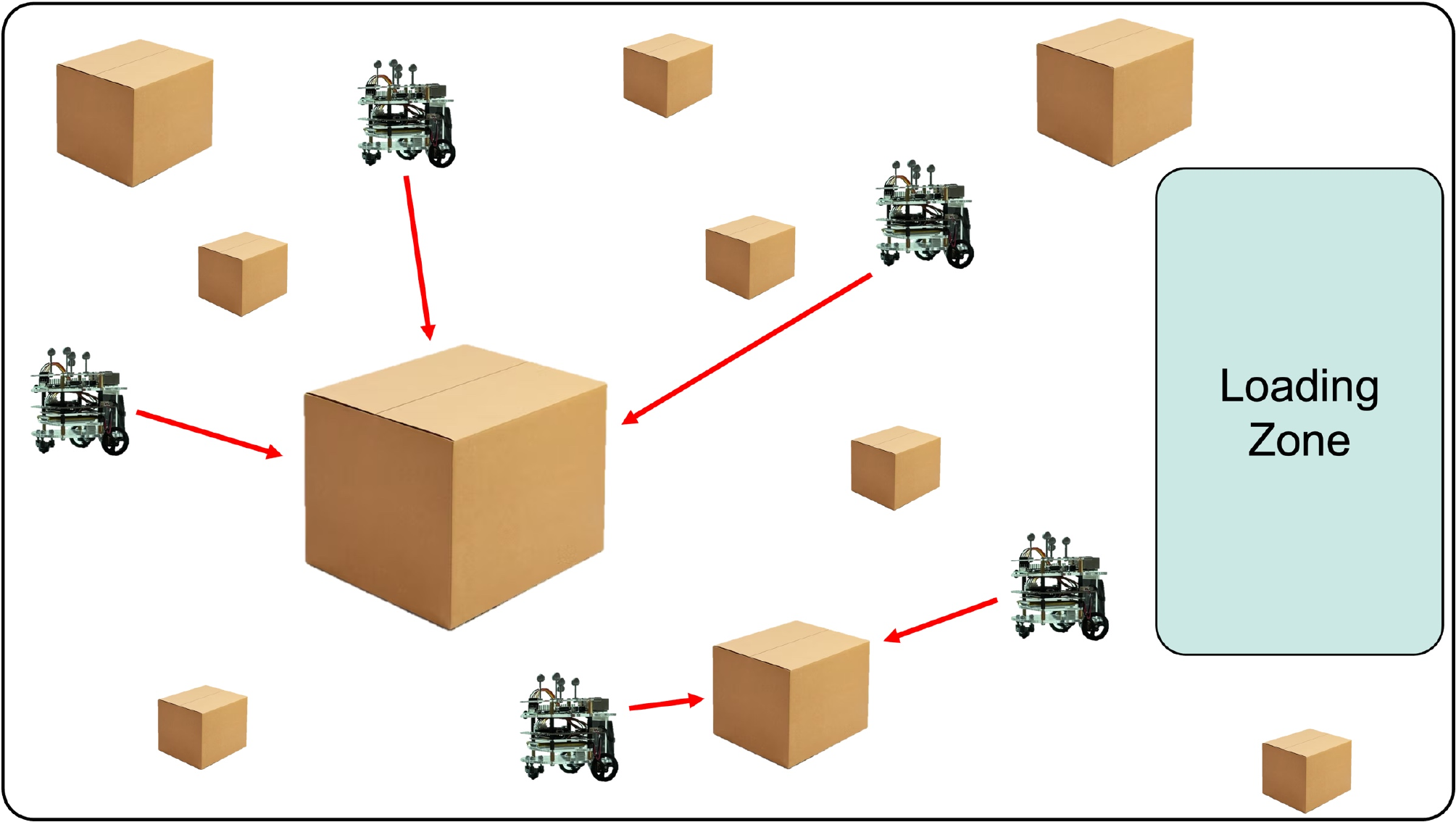}
        \label{fig:collab}
    }
    \caption{Illustrative examples of cooperation, coordination, capability complementarity, and collaboration in a simplified warehouse environment with homogeneous robots and three different package sizes; small packages can be moved by any single robot, medium packages can be moved by any two robots, and the large package can be moved by any three robots. In all cases, robots share a common goal of moving all packages from the warehouse to a loading zone (depicted by the green area), which is the basis of cooperative ``intent'' between robots. Red arrows represent the task assignment of each robot (that is, which package to move) and a crossed-out red arrow represents a "bad" assignment (that is, redundant assignment or infeasible task). In (a), robots randomly, or greedily, go to move packages without regard for the selection of other robots (that is, no coordination) and without taking joint action(s) to move larger packages (that is, no capability complementarity), which can lead to suboptimal assignments and exclude all packages that require more than one robot to transport. In (b), robots share information to determine what small box each robot must move, but do not take any joint action to move packages together (that is, no capability complementarity). In (c), robots randomly, or greedily, go to move packages (that is, no coordination); however, robots may form joint capabilities, should they happen to arise, and move packages together. In (d), robots share information to determine what box size each robot must move, or what arrangements in terms of capability complementarity are needed to clear the warehouse of all box sizes.}
    \label{fig:full_illustrative_example}
\end{figure*}

As an example, consider a simplified warehouse environment with homogeneous robots and packages of varying sizes that may require one or more robots to transport (see Figure~\ref{fig:coop}), where the \textit{shared goal} of all robots is to transport packages to a designated loading area. Each robot is then \textit{tasked} to grab packages and offload them in the loading zone. The common goal between robots provides a basis for cooperation, where, supposing that each robot selects a package at random (or through a greedy mechanism) to retrieve and transport it to a loading zone, and that the packages are all small enough to be moved by a single robot, then robots can behave independently while still contributing to the global objective of clearing the warehouse. Thus, even though such a scenario may be viewed as suboptimal with respect to the efficiency and capability of the entire system, the shared non-adversarial intent among robots to achieve this collective goal satisfies the definition of cooperation as presented in Definition~\ref{def:cooperation}. 

The warehouse could, in principle, also be cleared by robots with adversarial intentions, even if they receive shared rewards for completing the task. This survey focuses on robots with non-adversarial intentions toward one another while pursuing a shared goal or related tasks, as specified in Definition~\ref{def:cooperation}. Collaboration in the proposed taxonomy builds on cooperation and therefore inherits this non-adversarial scope. This is a choice about the relationships examined in the survey, rather than a claim that successful task completion requires non-adversarial intent.

Having established cooperation as the foundational layer of collective behavior, we now strengthen the behavioral requirement(s) by considering how robots, on top of shared intent and the absence of antagonism, should couple their actions to achieve more structured system behavior.

Coordination, as we define in Definition~\ref{def:coordination}, requires additional structure beyond shared intent. Robots must exchange information and/or follow rules that ensure their actions remain compatible with one another. This \textit{coupling} may arise from task dependencies, resource constraints, or the need to avoid conflict during execution. While coordination can improve team efficiency, it does not necessarily yield new capabilities that surpass those of any individual robot.

\begin{definition} \label{def:coordination}
    \textbf{Coordination} between two or more robots is the cooperative \textbf{information exchange} to plan their actions and/or decisions, in order to execute interdependent tasks more effectively than any single robot operating alone.
\end{definition}

For this taxonomy, information exchange is relevant when it is used to couple task decisions. Merely sensing another robot or receiving a message does not establish coordination. The information may be communicated explicitly or obtained through local observation, as described above; its role in planning interdependent actions is the distinguishing requirement.

Returning to the same warehouse environment, let's suppose that each robot is able to broadcast (that is, share information) which package it intends to retrieve based on proximity (see Figure~\ref{fig:coord}). This additional information exchange then prevents two (or more) robots from redundantly going to move the same small package or even blocking each other's paths. Robots can now use this information to assign themselves, or be assigned, to different packages, leading to a more efficient and potentially faster achievement of the shared goal. The task structure has not changed as each package is still small enough to be carried by a single robot, but the robots' actions are now \textit{coupled}, which is necessary in coordination. The decision made by one robot directly influences the decisions and/or actions available to others. 

While coordination plans independent robot actions, it does not require robots to jointly execute tasks that exceed their individual capabilities. This motivates the next category of interaction in the taxonomy presented in this article: capability complementarity.

\begin{definition} \label{def:capability_complementarity}
    \textbf{Capability Complementarity} between two or more robots is the cooperative use of their joint capabilities to execute tasks unattainable by any individual robot alone, thereby expanding the \textbf{joint action} set.
\end{definition}

The joint action set in Definition~\ref{def:capability_complementarity} is interpreted at the level of task execution: combined actuation, sensing, or other capabilities make an outcome attainable that is unavailable to each robot acting alone under the same task assumptions. This does not require introducing a new primitive command into an individual robot's controller. Communication can enable such an outcome when it connects complementary capabilities, but exchanging messages alone establishes neither their necessity nor the resulting joint capability. A temporal requirement must likewise be part of the same stated task for the individual and the team. Faster completion or satisfaction of a deadline alone is insufficient evidence; the description must identify which combined capability makes the required outcome attainable.

\begin{quote}\itshape
    Achieving collaboration in multi-robot systems relies on non-adversarial intention, information exchange, and joint action.
\end{quote}

Capability complementarity (Definition~\ref{def:capability_complementarity}) concerns what robots can accomplish through their combined capabilities, while coordination (Definition~\ref{def:coordination}) concerns how information exchange couples their decisions. The distinction is illustrated by joint payload transport in \cite{zhang2020decentralized} and the integration of observations beyond an individual robot's sensing range in \cite{Zhou2022}. Coordinated assignment of small packages need not create either kind of joint capability. Conversely, Figure~\ref{fig:capab_comp} isolates capability complementarity through an unplanned encounter in which robots move a larger package together. This illustrative case separates the criteria rather than prescribing a strategy for reliably arranging joint actions. The warehouse example also shows that homogeneous robots can exhibit complementarity: their identical individual capabilities can still be insufficient for a task requiring their combined effort. The survey focuses on the intersection of coordination and capability complementarity within cooperation, labeled collaboration in Figures~\ref{fig:venn_diagram} and~\ref{fig:collab}.

\begin{definition} \label{def:collab}
    \textbf{Collaboration} is cooperation, coordination, \textit{and} capability complementarity, between two or more robots.
\end{definition}

In the proposed taxonomy, coordination and capability complementarity are defined as cooperative relationships, so their intersection already inherits cooperation. Definition~\ref{def:collab} states cooperation explicitly to keep the shared, non-adversarial intent visible alongside information exchange and joint capability. Within this taxonomy, all three conditions are necessary and together sufficient for the collaboration classification; they are not three logically independent requirements.

In Figure~\ref{fig:collab}, robots plan the joint transport of medium and large packages that cannot be moved individually. The example therefore combines coordination with capability complementarity. Figure~\ref{fig:capab_comp} illustrates that a larger package may also be moved through an unplanned encounter; such a case already uses capability complementarity. The restriction illustrated by coordination alone in Figure~\ref{fig:coord} is narrower: assigning individually executed transport actions does not provide the combined capability required by the larger packages.

To apply Definition~\ref{def:collab}, we identify the shared, non-adversarial objective, the information exchange coupling robot decisions, and the task outcome enabled by combined capabilities. The last criterion is assessed under the same task constraints for individual and joint execution, as illustrated by the package sizes in Figure~\ref{fig:full_illustrative_example}. Improved efficiency alone is insufficient to establish capability complementarity and hence collaboration. These criteria guide the organizational and methodological comparisons that follow.

Redundancy illustrates why the task requirements matter when applying these criteria. When any robot can independently satisfy the complete task requirements, deploying additional robots for reliability does not by itself establish capability complementarity. Such execution is cooperative when robots share non-adversarial intent, and coordinated when information exchange couples their decisions, for example in allocating work or arranging a handover. If the task instead explicitly requires an outcome under failures that no individual robot can satisfy alone, capability complementarity depends on whether the robots' combined capabilities make that outcome attainable. Collaboration then also requires cooperation and coordination. An execution policy can determine how a joint capability is realized, but the use of multiple robots alone does not establish its necessity; the same task requirements must be used when assessing individual and joint feasibility.

For a specified task, its constraints, and a team of two or more robots, the collaboration definition proposed in this survey (Definition~\ref{def:collab}) can be expressed through the following conditions:
\begin{enumerate}
    \item The robots have non-adversarial intentions to contribute toward a shared goal or related tasks (cooperation, Definition~\ref{def:cooperation}).
    \item They use cooperative information exchange to plan their actions and/or decisions to execute interdependent tasks more effectively than any single robot operating alone (coordination, Definition~\ref{def:coordination}).
    \item They cooperatively use their combined capabilities to execute a task unattainable by any individual robot alone under the same task assumptions, thereby expanding the joint action set (capability complementarity, Definition~\ref{def:capability_complementarity}).
\end{enumerate}
These conditions are individually necessary and jointly sufficient for collaboration within the proposed taxonomy. They summarize the definitions adopted in this survey and provide a basis for interpreting the surveyed works under those definitions.

Together, these definitions provide a structured hierarchy of collective behaviors in multi-robot systems, distinguishing non-adversarial shared intent (cooperation), information sharing (coordination), and joint action(s) (capability complementarity). We now use this taxonomy to present a background on the cooperative (Definition~\ref{def:cooperation}) and coordinated (Definition~\ref{def:coordination}) multi-robot literature.

\section{Background on Cooperation and Coordination}\label{sec:background}

In the control literature emphasized in this background section, multi-robot systems are commonly modeled within the framework of networked systems, where each robot is represented as a node in a graph and edges encode inter-agent relationships, such as communication and sensing \cite{Mesbahi10:Graph}. Within this setting, a global objective, shared among the robots, is achieved through interaction rules, and a rich body of work exists, focusing on goals such as consensus, formation, and coverage. Foundational works such as \cite{olfati2004consensus, Zavlanos11:Graph} provide general graph-theoretic and systems-theoretic tools for analyzing these behaviors. Complementary surveys on consensus and distributed coordination \cite{chen2019control, li2019survey, qin2016recent} formalize what it means for agents to work toward a shared objective. In this section, we focus on classical coordination (Definition~\ref{def:coordination}) problems involving cooperative (Definition~\ref{def:cooperation}) robots. The following discussion focuses on non-adversarial interactions and the mechanisms used to couple robot decisions. These mechanisms provide background for the task-dependent collaboration criteria introduced above.

In much of the literature on multi-robot systems, cooperation and coordination are often treated implicitly, being captured through modeling assumptions in problem formulations, without being defined explicitly. In consensus problems, for example, the robots iteratively update their states (such as, positions, headings) based on their neighbors' information so that all agents asymptotically "agree", that is, converge, on a common value \cite{olfati2004consensus, ren2005survey}. Consensus has become a typical example of a model for coordinated behavior (Definition~\ref{def:coordination}) between cooperative robots (Definition~\ref{def:cooperation}), underlying flocking, rendezvous, distributed estimation, and distributed optimization. Surveys such as \cite{li2019survey, qin2016recent, amirkhani2022consensus} analyze variants including finite-time consensus, event-triggered consensus, and consensus under communication constraints, and clarify how local interaction rules give rise to desired group-level behavior.

Formation control also provides a classical instance of coordinated multi-robot behavior (Definition~\ref{def:coordination}), where the goal is to maintain prescribed relative configurations or geometric patterns among robots. Graph rigidity and distributed formation stabilization show how inter-agent distance constraints and graph-theoretic properties determine whether a desired shape can be maintained \cite{olfati2002graph}. Other surveys, such as \cite{oh2015survey, liu2024survey}, review approaches ranging from leader-follower and virtual-structure methods to behavior-based and optimization-based schemes. In these frameworks, coordination is achieved through coupled relative-position or relative-velocity feedback laws that ensure alignment of individual actions with a shared formation objective.

As another example of multi-robot coordination (Definition~\ref{def:coordination}), coverage control focuses on spatial allocation of robots, that is, positioning a group of robots to optimally cover a domain of interest. A common formulation is the Voronoi tessellation-based coverage control \cite{Cortes04:Coverage}, where robots perform gradient descent with respect to a cost function to achieve a team-level configuration, that is, a centroidal Voronoi tessellation. This formulation has been extended to scenarios with, for example, non-convex environments \cite{breitenmoser2010voronoi}, heterogeneous sensing capabilities \cite{mahboubi2016distributed}, and time-varying densities \cite{lin2026disentangled}, and is closely tied to distributed optimization and geometric partitioning of the workspace, such as, \cite{nedic2009distributed, schwager2009optimal, rekleitis2004limited, hazon2005redundancy, karapetyan2017efficient, kapoutsis2017darp}.

Coordination also arises through distributed optimization and distributed decision-making. The coverage and allocation methods discussed above include distributed optimization mechanisms \cite{nedic2009distributed, schwager2009optimal}. In \cite{kloock2023synchronization}, local planning alternates with synchronization of the plans exchanged among agents, addressing consistency in distributed decisions. The tracking model predictive control (MPC) approach in \cite{koehler2024distributed} uses locally optimized artificial references communicated to neighboring agents; a coupling cost connects these reference choices to the cooperative objective. The coalitional tracking approach in \cite{chanfreut2022tracking} organizes local controllers into changing coalitions whose members coordinate their inputs for target-set tracking. These approaches extend information-dependent coordination from agreement on states or geometric configurations to interdependent planning and control decisions.

Learning-based coordination is illustrated by decentralized collective construction in \cite{sartoretti2019distributed}, where robots learn coordinated block-placement policies under structural constraints. On the other hand, collective transport also illustrates coordination through local sensing and physical feedback. The stochastic attachment and detachment policies in \cite{wilson2014ant} allocate robots around loads, drawing on ant transport behavior. The study in \cite{wilson2018towing} uses decentralized robot towing controllers to examine how local force feedback and differences among transporters affect collective motion.

\begin{table*}[!t]
    \centering
    \small
    \caption{Terminology and organizing perspectives in the reviewed literature, with their interpretation under this survey's taxonomy.}
    \begin{tabular}{p{0.27\textwidth}p{0.29\textwidth}p{0.33\textwidth}}
        \hline
        \textbf{Term or perspective in the reviewed literature} & \textbf{Emphasis described in this survey} & \textbf{Reading under the present taxonomy} \\
        \hline
        Coordinated control \cite{Cortes17:Coordinated, yan2013survey} & Information-dependent motion, task decisions, and interaction structure. & Coordination when information couples interdependent decisions under cooperative intent. \\
        Cooperative heterogeneous systems \cite{Rizk20:Cooperative} & Shared tasks and differing robot capabilities. & Heterogeneity alone does not establish complementarity; the required joint capability is task-dependent. \\
        Resilience and industrial collaboration surveys \cite{prorok2021beyond, menebroker2025multi} & Reliability-related behavior and industrial applications, respectively. & These organizing perspectives may overlap with all three relationships; they are not separate interaction categories here. \\
        Cooperative and noncooperative games \cite{marden2018game, emery2005game} & Agreements, coalitions, and strategic decisions. & Game-theoretic terminology is distinct from shared non-adversarial intent; the task and interactions determine the classification. \\
        Collaborative learning and CTDE \cite{zhang2018fully, Orr23:Multi_Agent} & Shared information, learning objectives, or training processes. & These can support coordination; collaboration additionally requires a task outcome enabled by combined capabilities. \\
        \hline
    \end{tabular}
    \label{tab:terminology_reading}
\end{table*}

Existing surveys organize this literature through several complementary perspectives. Coordinated control in \cite{Cortes17:Coordinated} emphasizes motion tasks such as rendezvous, formation, and coverage, while \cite{yan2013survey} considers communication models, task structures, and scalability. Market-based coordination \cite{dias2006market} and cooperative heterogeneous systems \cite{Rizk20:Cooperative} provide further views of allocation and team composition. The resilience taxonomy in \cite{prorok2021beyond} centers on resilient multi-robot systems, whereas \cite{menebroker2025multi} surveys mobile-robot collaboration in industrial applications. The present survey instead uses shared intent, information-coupled decisions, and task-dependent joint capabilities as a common basis for examining cooperation, coordination, and collaboration across applications.

Building on these foundations, cooperation, coordination, and collaboration are treated as related yet conceptually distinct forms of multi-robot interaction according to the taxonomy introduced in the previous section. This taxonomy provides the basis for examining how multi-robot systems are organized and structured to enable collaboration, as per Definition~\ref{def:collab}, in both theory and practice.

Table~\ref{tab:terminology_reading} summarizes how terminology and organizing perspectives already encountered in this review relate to the proposed taxonomy. The final column gives this survey's interpretation of the described mechanisms; it does not attribute new definitions to the cited authors.

\section{Multi-Robot Collaboration}\label{sec:collaboration}

We next examine organizational structures and frameworks that support collaborative robot tasks. The discussion includes both task examples with identified joint capabilities and enabling control, learning, or interaction methods. The latter are considered for the mechanisms they contribute; applying the collaboration classification also requires the task-dependent evidence specified in the taxonomy section.

\subsection{Organizational Structures}\label{sec:organization}

\begin{figure*}[!b]
    \centering
    \subfloat[Centralized]{
        \includegraphics[width=0.485\textwidth]{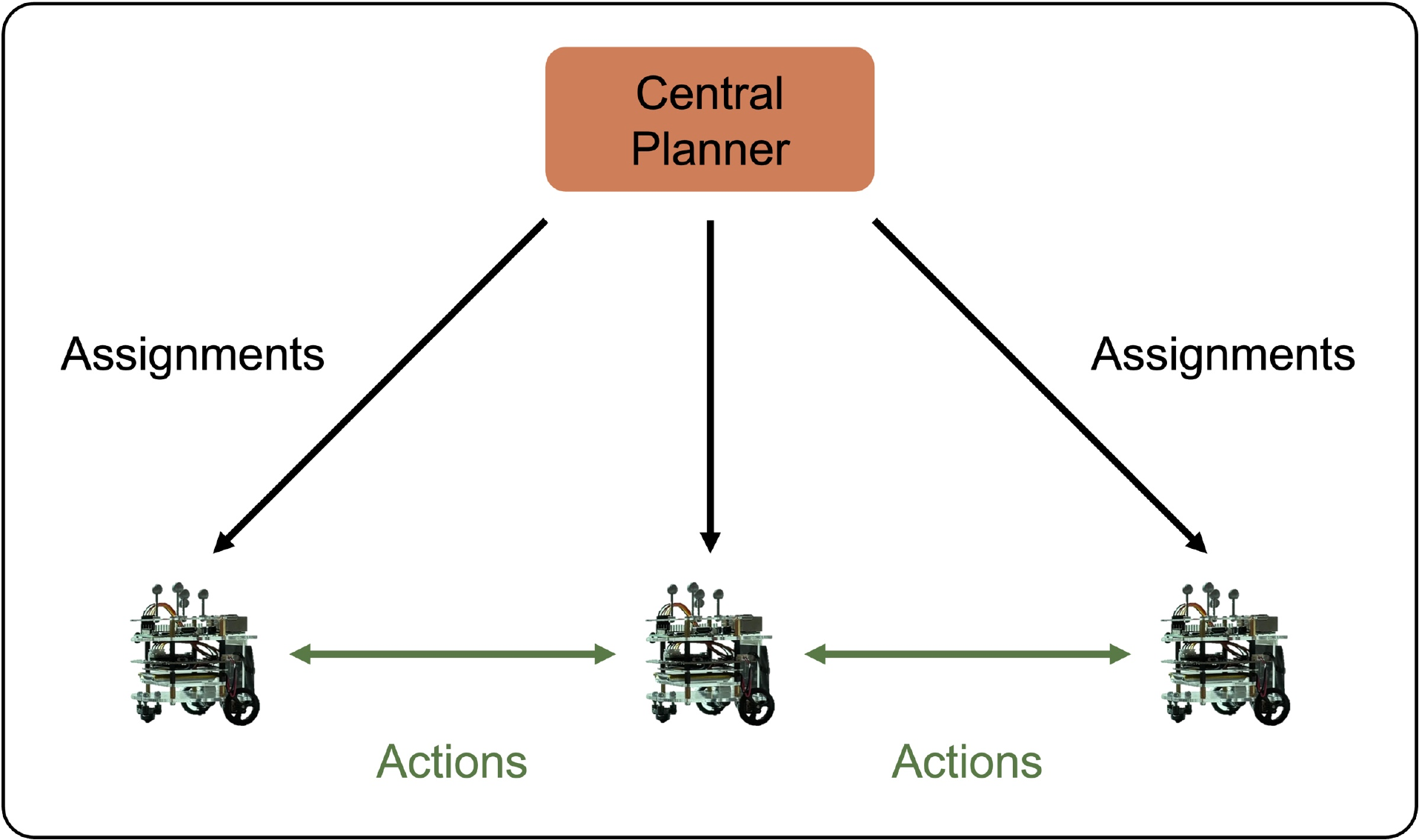}
        \label{fig:centralized}
    }
    \hfill
    \subfloat[Decentralized]{
        \includegraphics[width=0.485\textwidth]{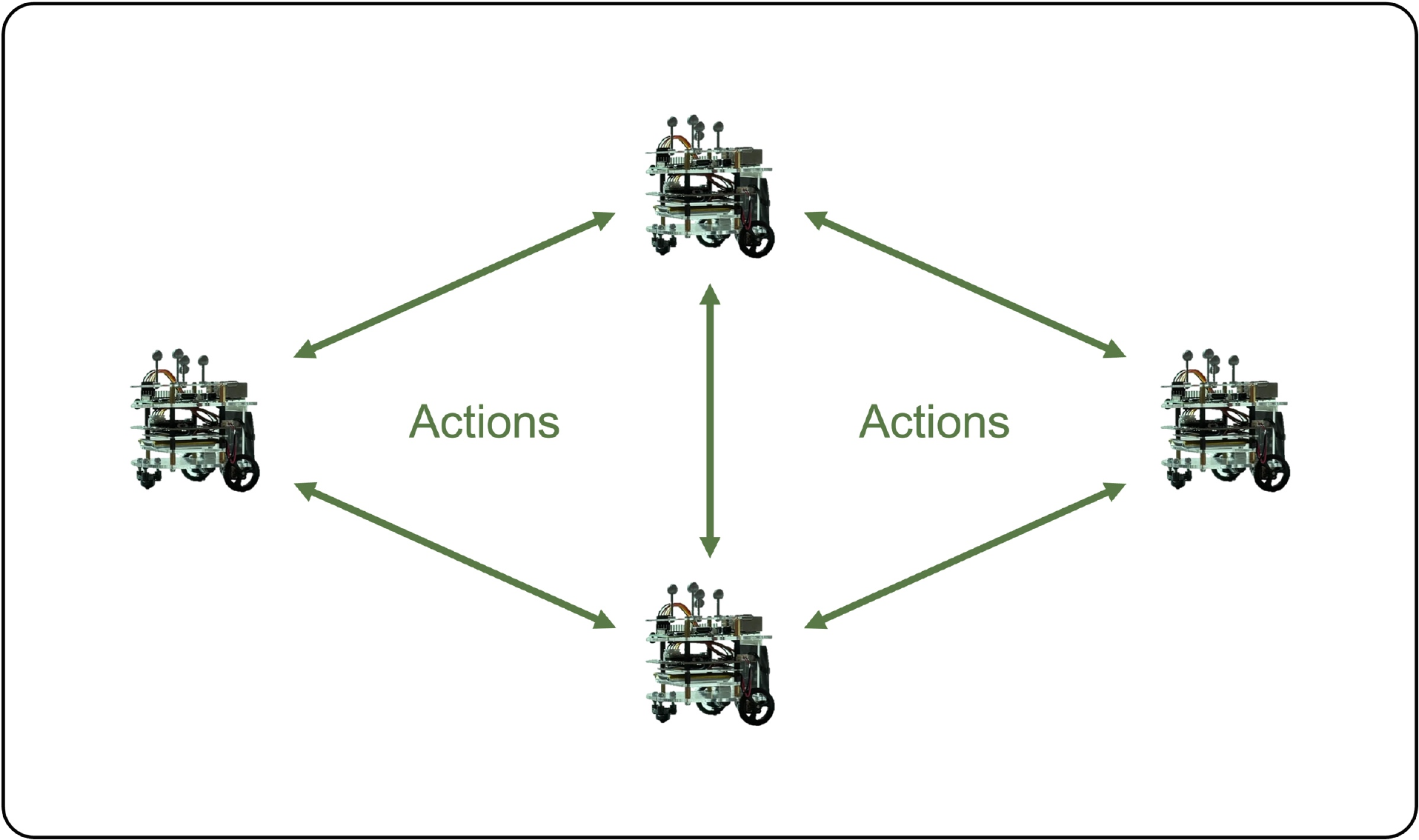}
        \label{fig:decentralized}
    }
    \hfill
    \subfloat[Hierarchical]{
        \includegraphics[width=0.485\textwidth]{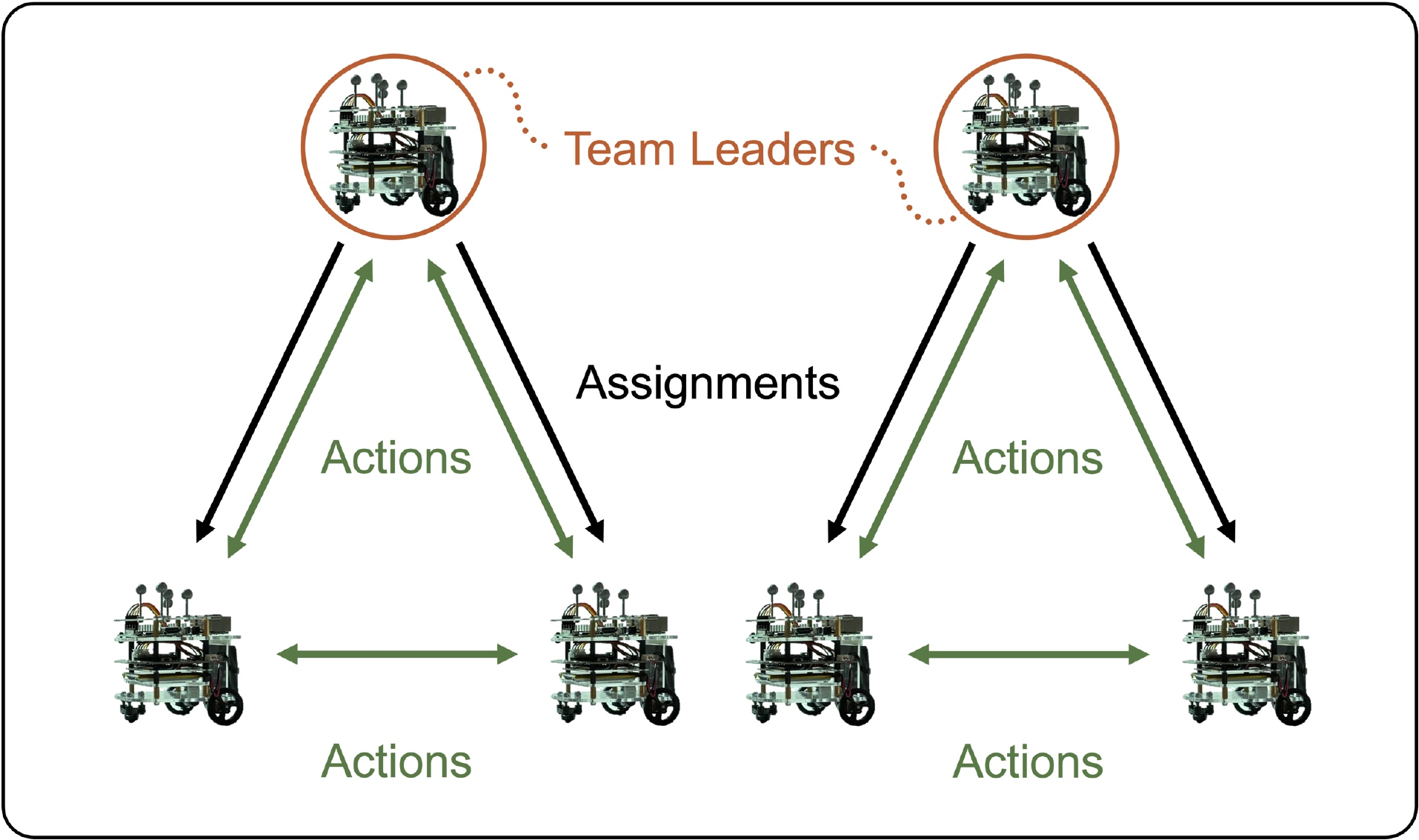}
        \label{fig:hierarchical}
    }
    \caption{Illustrative example of organizational structures for collaborative multi-robot systems. In (a), a central planner assigns tasks to each robot, and robots engage in collaborative actions only when directed by the controller. In (b), robots independently decide when to collaborate often through communicating requests for assistance. In (c), the robots are divided into teams with one robot acting as team leader. The leader instructs the other robots on when and how to take collaborative actions.}  
    \label{fig:system_org}
\end{figure*}

Centralized, decentralized, and hierarchical organizational structures are well established in multi-robot systems \cite{yan2013survey, jawhar2018networking, scattolini2009architectures}. Their relevance here is that collaboration can arise under each of these arrangements; it does not depend on a particular organizational structure. The distinction concerns who decides if, when, and with whom robots collaborate, while the collaboration criteria concern how robots combine their intent, information, and capabilities to accomplish a task. Figure~\ref{fig:system_org} illustrates these arrangements. The following discussion highlights how each can support team formation and joint execution through representative examples from the surveyed literature.

\subsubsection{Centralized}

A centralized planner assigns tasks and roles using a global view of the system \cite{yan2013survey}. This can help it consider the needs and capabilities of multiple robots when deciding which tasks require a team and which robots should participate. Individual robots then execute the assigned actions. The arrangement concentrates computation and communication at the planner, whose failure can affect the team; these demands can also become a constraint as the number of robots grows \cite{jawhar2018networking, yan2013survey}.

Multi-robot task allocation (MRTA) provides a setting for these decisions, using optimization, auctions, or game-theoretic methods \cite{Gerkey04:A_Formal, badreldin2013comparative}. The centralized queue in \cite{nguyen2024scalable} matches collaboration requests with available robots according to a selected strategy, such as first in, first out or a matching algorithm. These pairings connect robots needing assistance with partners whose capabilities expand their accessible operating regions.

In \cite{liu2024coherent}, a large language model (LLM) plans for a heterogeneous team consisting of a robotic arm, a quadruped robot, and a quadrotor. It decomposes a human instruction into subtasks and assigns roles according to robot capabilities, including tasks requiring joint manipulation.

\subsubsection{Decentralized}

Decentralized approaches place collaboration decisions with individual robots, as in \cite{lin2025heterogeneous}. Local decision-making removes dependence on a single planner and can support scalability \cite{yan2013survey, jawhar2018networking}. Each robot must decide whether and how to participate using the information available to it, while the participating robots must reach compatible decisions. Communication, sensing, and negotiation therefore play a role in arranging the joint activity, as reflected in the terminology and examples below.

In this survey, ``decentralized'' is used in the broad sense of non-centralized decision-making: individual robots determine their actions using locally available information. This information may come from wireless communication, local sensing of neighboring robots and the environment, or a combination of these sources. These information pathways can support task decisions and safety-related actions, such as collision avoidance. In networked-control terminology, decentralized control without inter-agent communication is often distinguished from distributed control with inter-agent communication. Our architectural category encompasses both, including sensing-mediated interactions; it describes where decisions are made, independently of how the information used in those decisions is obtained.

Local team formation can be organized through negotiation. In \cite{shan2024distributed}, robots bid for time-constrained transport tasks whose size or weight requires joint manipulation, and form teams through decentralized negotiation. Centralized training with decentralized execution (CTDE) instead uses shared global information to train policies that robots execute independently \cite{Orr23:Multi_Agent}. For tasks requiring complementary capabilities, this separates the information available during learning from the decisions robots must make during joint execution. The learned coalition-selection methods discussed in ``Learning-Based Frameworks'' likewise use decentralized task decisions during execution.

\subsubsection{Hierarchical}

Hierarchical structures divide decision-making responsibilities across levels: leaders determine tasks or team arrangements, while robots at lower levels execute the assigned actions \cite{yan2013survey}. This can accommodate differences in task knowledge or computational resources among agents. For collaborative tasks, the hierarchy connects decisions about if, when, and how a team should act with the robots that carry out the joint activity. The coalition and supervisory examples below illustrate this division of responsibility.

Coalition formation can also distribute authority across leaders and followers. In \cite{zhang2024coalition}, leader robots organize the resources needed by a task, while followers choose coalitions using an altruistic preference. The resulting teams pool resource types and quantities that individual robots cannot supply for those tasks. This division of responsibilities connects hierarchical organization with game-theoretic team formation.

Hierarchical structures also arise in HSI frameworks when a human or supervisory agent provides high-level mission guidance, as in \cite{hussein2022characterization, divband2021designing}. Supervisory decisions and local robot execution occupy different levels of the hierarchy.

\subsection{Framework Inspirations and Tools}\label{sec:frameworks}

\begin{figure*}[!b]
    \centering
    \includegraphics[width=0.75\columnwidth]{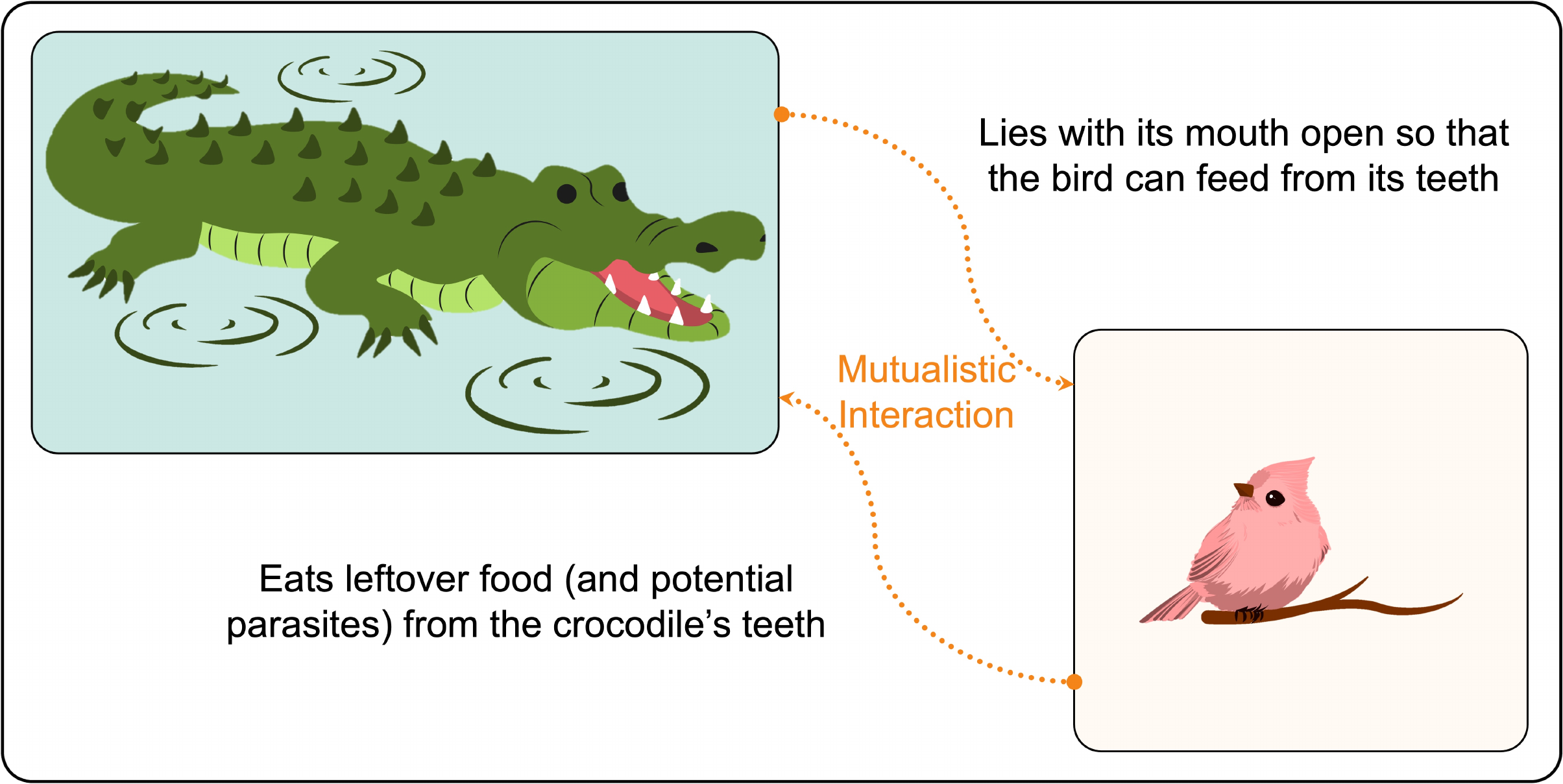}
    \caption{
Illustrative example of mutualism between two species, the Nile crocodile and the Egyptian plover \cite{b1961scientific, molles1999ecology}. The depicted interaction combines oral cleaning for the crocodile with food acquisition for the bird.
}
    \label{fig:mutualism_example}
\end{figure*}

The organizational structures described above determine where decisions about collaboration are made. This section examines the modeling and design approaches through which robots combine their capabilities to accomplish tasks beyond those achievable individually. We first consider ecology-inspired, human-swarm interaction (HSI), game-theoretic, and learning-based frameworks, examining their contributions to the formation and execution of collaborative arrangements.

The choice of these four categories is motivated by the specific mechanisms they offer for supporting collaboration. Ecology-inspired frameworks model relationships through which robots with complementary capabilities can expand their feasible actions or accessible operating regions \cite{Nguyen23:Mutualistic, lin2025heterogeneous}. HSI frameworks connect supervisory guidance with swarm feedback and autonomy allocation, providing mechanisms for directing robot teams engaged in collaborative tasks \cite{hussein2022characterization, divband2021designing}. Game-theoretic frameworks address the formation of coalitions and the allocation of utility among robots whose combined resources or capabilities are needed for a task \cite{zhang2024coalition, shan2024distributed}. Learning-based frameworks provide policies and shared representations for realizing joint capabilities, including collective object transport and perception from complementary observations \cite{zhang2020decentralized, Zhou2022}. The four categories do not span the entire collaborative multi-robot literature. The final subsection, "Additional Collaboration Frameworks," extends the discussion to collaboration in other domains, including task allocation, joint manipulation, and the integration of heterogeneous robot platforms.

\subsubsection{Ecology-Inspired Frameworks}

Natural systems and the environment have long served as a rich source of inspiration for modeling complex systems; understanding their structure and dynamics has influenced a substantial body of prior work \cite{bonabeau1999swarm, benyus1997biomimicry, bar2005biomimetics, vincent2006biomimetics}. Ecology is the study of interactions between organisms and their environment \cite{ricklefs2000ecology}, and collaborations can also be conceptualized as mutualisms (see Figure~\ref{fig:mutualism_example})---jointly beneficial interactions between members of different species \cite{Pauli14:A_Syndrome}. In multi-agent robotics, ecological principles have informed constraint-driven control methods and frameworks for combining robot capabilities \cite{egerstedt2018robot, notomista2019constraint, Nguyen25:Mutualisms}.

The authors in \cite{Nguyen23:Mutualistic} introduce a mutualistic collaboration framework for heterogeneous multi-robot systems, in which robots with different mobility capabilities form pairwise collaborative arrangements based on the composition of control barrier functions (CBFs). Through this composition, each robot's safe operating region can expand, enabling robots to accomplish tasks and reach states that could not be achieved alone. Similarly, a multi-robot collaboration framework based on mutualisms was introduced in \cite{Nguyen25:Mutualisms}, which examined how the composition of the landscape can dictate when collaborative arrangements between different types of robots are preferred and when they are not. Others have proposed a more general symbiotic framework that considers mutualism, commensalism, and parasitism in a multi-agent reinforcement learning (MARL) setting \cite{Niu25:Investigating}. Here, we consider its relevance to non-adversarial joint tasks; interactions involving deliberate harm or interference fall outside the proposed taxonomy's cooperative scope.

Ecology has also inspired the constraint-driven control synthesis framework in \cite{egerstedt2018robot, notomista2019constraint}, where robots (organisms) tend to minimize their control efforts, subject to environmental constraints (such as, collision avoidance and battery capacity maintenance). This ecology-inspired framework \cite{notomista2019constraint} is formulated as an optimization problem that utilizes control Lyapunov functions (CLFs) and CBFs \cite{ames2019control}. These control tools can serve different task requirements. In the formation experiment in \cite{notomista2019constraint}, decentralized control uses inter-robot distances to assemble a hexagon. When the prescribed spatial configuration is itself the required output, its realization combines robot positions that no individual robot can occupy simultaneously, identifying the joint capability in that task.

Building upon \cite{notomista2019constraint}, a heterogeneous multi-robot collaboration framework is proposed in \cite{lin2025heterogeneous} by explicitly characterizing the environment with which robots (organisms) interact by a set of partial differential equations. Collaborative behaviors are achieved through principled compositions of CBFs and CLFs, adapted from \cite{glotfelter2017nonsmooth}. For example, in terms of collaboration in mobility, inspired by \cite{kim2022heterogeneous}, some agents' safe operating regions in the state space can expand through interactions with other agents, enabling access to regions that would otherwise be unreachable. In terms of collaboration in sensing, inspired by \cite{santos2018coverage,santos2018coveragelimited}, heterogeneous robots equipped with complementary sensing modalities collaborate to monitor time-varying environmental phenomena that are inaccessible to some individuals.

The transport study in \cite{alkilabi2016transport} considers homogeneous e-puck robots transporting objects too heavy for a single robot. Each robot uses an evolved neural controller with displacement feedback to relate pushing actions to object motion. Here, the required combination of pushing capabilities supplies a concrete link to the warehouse example, while the feedback supports coordination through the shared physical task.

These examples illustrate different roles for ecological inspiration: composing constraints to enable joint mobility and sensing, shaping interactions through rewards, and coordinating combined pushing through local feedback. They connect the representation of joint capabilities to the stability, safety, and performance questions discussed in the final section.
\subsubsection{Human-Swarm Interaction Frameworks}

Human-swarm interaction (HSI) connects human mission guidance with the sensing, motion, and manipulation of robotic teams \cite{kolling2015human}. Human input can enable joint execution by specifying the desired outcome and adapting it to the task context, while robot controllers translate that guidance into interdependent actions. This supervisory and interaction role motivates the HSI frameworks considered here. Within the proposed taxonomy, Definition~\ref{def:collab} applies to the combined robot capabilities used to realize the outcome; the human contribution concerns how those capabilities are directed and brought into use.

Mission, interaction, complexity, automation, and human (MICAH), introduced in \cite{hussein2022characterization}, provides a conceptual framework for adjusting autonomy and human involvement using indicators in these five categories. Its illustrative search-and-rescue scenario combines a human-operated aerial robot with ground robots. The aerial platform offers better sensing for assessing victim presence, while ground robots approach victims and enter areas inaccessible to the aerial platform. Aerial guidance informs ground activity. The integrated assessment and retrieval scenario thus brings together sensing and access capabilities unavailable on either platform alone; MICAH addresses how supervisory involvement can adapt during that activity.

Distributed perception combines observations across a robot team. In \cite{divband2021designing}, robots exchange belief and confidence maps through local interactions, extending perception beyond their individual sensing ranges. The shared confidence maps guide exploration toward less-observed regions, coupling motion decisions to observations collected across the team. The interface allows an operator to inspect these maps and direct exploration through nearby robots.

Virtual reality (VR)-based coverage control connects human mission priorities with spatial deployment \cite{figueiredo2019voronoi}. The operator changes density functions, obstacles, and robot parameters, giving the team a coverage objective that reflects the desired activity. The robots adapt their configuration to this guidance. Such an interface can enable collaborative sensing by directing the deployment of a team's combined sensing capabilities when the mission requires observations beyond those available to an individual robot.

Gesture and sketch interfaces let the human specify the spatial outcome that the swarm should realize. SwarmPaint \cite{serpiva2021swarmpaint} uses learned gesture recognition to translate human input into trajectories and formations; its simulated formation-targeting task involves positioning and rotating a line or triangle while adjusting inter-robot separation. Freehand sketching in \cite{karam2026freehand} extracts formation points from a drawing, constructs a rigid graph, and assigns robots to positions for distributed formation execution. Human input determines the desired geometry, and the robots jointly realize it through their simultaneous presence at distinct positions. Within the proposed taxonomy, this specified geometric output supplies the joint capability, since one robot cannot occupy those separated positions simultaneously.

Augmented reality (AR) can connect human guidance with complementary manufacturing operations. The framework in \cite{li2022ar} combines AR and a digital twin with robot control. Its dual-robot disassembly demonstration assigns predefined operations to a KUKA robot while a human teleoperates a mobile UR robot for assisting actions such as dragging and pick-and-place. Human guidance brings autonomous processing and mobile handling into a shared workflow and adapts assistance to the task. Balancing supervisory guidance, robot autonomy, and task dependencies connects this setting to the research directions discussed in the final section.

\subsubsection{Game Theoretic Frameworks}

Game theory models strategic interactions among decision-makers. Its terms ``cooperative'' and ``noncooperative'' concern the formulation of agreements and strategic decisions, whereas cooperation in Definition~\ref{def:cooperation} concerns shared, non-adversarial intent. A noncooperative game formulation can therefore organize robots pursuing a shared task without implying adversarial behavior \cite{marden2018game, emery2005game}. The path-planning method in \cite{qiu2025multi} combines game-theoretic decisions with particle swarm optimization to coordinate routes under motion and collision constraints. The learning-based MPC approach in \cite{le2024multi} couples robot navigation decisions in crowds. These methods provide tools for coordinating motion during joint tasks.

Resource-constrained task allocation in \cite{liu2025game} uses distributed coalition decisions to combine the resources supplied by participating robots. For tasks whose required resource combination is unavailable to any individual robot, this pooling supplies the joint capability, while the game organizes the robots' contributions. Cooperative game theory provides tools for modeling binding agreements and coalitions. When the task requires pooled capabilities, these tools can help organize their joint use. Frameworks such as characteristic-function formulations and transferable-utility (TU) games \cite{branzei2008models,elkind2016cooperative} quantify how groups of robots (that is, coalitions) can pool heterogeneous resources, sensing, or actuation capabilities to expand the collective action set. Solution concepts such as the core, Shapley value, and nucleolus address the division of coalition utility. This allocation question is distinct from whether a coalition has capabilities required to perform a task that its members cannot perform individually.

Coalitional control in \cite{fele2017coalitional, fele2018coalitional} provides methods for selecting which agents coordinate their decisions and how coalition benefits are allocated. The examples in these works concern energy trading and power-system control, respectively. Their relevance here is methodological: coalition negotiation and coupled control provide tools for organizing joint decisions when applied to robot tasks. For framework selection, the distinctive role of these methods is to connect coalition membership and utility to the decision to undertake a joint task. The distributed transport example in \cite{shan2024distributed} makes this connection concrete: negotiation arranges a team for a payload that requires multiple robots. Relating coalition decisions to task feasibility motivates the modeling questions considered in the final section.

\subsubsection{Learning-Based Frameworks}

Learning-based frameworks use policies and shared representations to support multi-robot coordination and collaboration \cite{wu2024state, Orr23:Multi_Agent}. The following examples examine their roles in joint action, team formation, perception, and task planning.

In multi-agent reinforcement learning (MARL), one robot's actions affect the learning dynamics and experience of others \cite{Tan1993, Whitehead1991}. In object transport, the task description in \cite{zhang2020decentralized} identifies a shared payload requiring synchronized force application and motion across robots, which cannot be moved by one robot under the stated constraints. Its decentralized learned policies act on that required joint physical capability.

Learning can also organize coalition formation and task assignment. In \cite{dai2024dynamic}, task requirements specify how many robots must act together, motivated by shared lifting; in \cite{dai2025heterogeneous}, teams must collectively supply the skills required to initiate a task. Their learned policies use information about other agents and task requirements to select subsequent tasks and form suitable coalitions. The relevant joint capability is the required combination of robot effort or skills, rather than a reduction in completion time alone. Both methods support decentralized task decisions. The leader-follower mechanism in \cite{dai2024dynamic} guides training; the heterogeneous method in \cite{dai2025heterogeneous} instead constrains action selection and decision ordering during training.

Learning can also guide the exchange of robots between teams with changing capability requirements. In \cite{karam2026altruistic}, Hamilton's rule motivates transfers by relating a receiving team's benefit to a donating team's cost. A graph neural network trained centrally uses neighboring teams' information to propose transfers during decentralized execution. In the firefighting task, robots have either sensing or water-carrying capabilities. Their combination supports detection and suppression that neither robot type can perform alone, while local feasibility checks preserve the required sensing presence as teams exchange robots.

Supervised learning can support distributed perception beyond an individual robot's sensing range. In \cite{Zhou2022}, a graph neural network (GNN) fuses encoded observations exchanged among neighboring robots. Integrating these complementary observations gives each robot a representation of its surroundings beyond its local view and expands the information available for decision-making.

Complementary viewpoints also support joint perception across robot platforms. In \cite{Miller22:Stronger}, learned semantic representations connect an aerial map with ground-level observations. Ground robots use the shared map to localize, select inspection targets, and plan paths through distributed onboard decisions. In the region-investigation task, aerial mapping and ground-level inspection combine viewpoints and access capabilities that neither configured platform supplies alone. The shared semantic representation connects these capabilities to navigation and inspection.

Foundation-model-based frameworks add language-mediated task planning to the learning tools considered here. In RoCo \cite{RoCo2024}, robots use large language models (LLMs) to discuss task strategies and generate sub-task plans and paths that incorporate high-level semantics and environmental feedback. This differs from the centralized example in \cite{liu2024coherent}, where one LLM decomposes a human instruction and assigns roles according to robot capabilities. The distinctive contribution of these examples is the connection between semantic instructions and the assignment or negotiation of complementary roles. In RoCo's sweeping example, pushing debris and collecting it provide the concrete joint manipulation capability. The role of the LLM is to support reasoning about and arranging those actions.

Taken together, the learning examples address different components of collaboration: RL produces task-execution policies, supervised learning fuses complementary observations, and LLMs support task reasoning and role planning. This comparison motivates the questions developed in the final section about learning when to collaborate, with whom, and under partial observability.

\subsubsection{Additional Collaboration Frameworks}\label{sec:additional_frameworks}

Collaboration can also be organized directly through task planning and the integration of robot platforms. The following examples connect task-planning and integration mechanisms with joint manipulation and sensing beyond the four modeling perspectives discussed above.

Task allocation provides a planning mechanism for determining which robots participate in a task and whether its execution requires a team. The centralized task-allocation method in \cite{martin2026} addresses the assignment problem by repeatedly decomposing it into smaller subproblems. Its collaborative task settings include combinations of robot capabilities that enable tasks unavailable to individuals. For these tasks, allocation determines a feasible team whose combined capabilities meet the task requirements.

Joint manipulation illustrates how collaboration can be embedded in the relationship between task planning and physical execution. In \cite{hekmatfar2014cooperative}, multiple mobile manipulators transport a shared object using a hierarchical framework. High-level planning coordinates task allocation and motion sequencing, while lower-level controllers execute synchronized manipulation actions. The payload cannot be transported by an individual robot under the task formulation, making combined physical actuation necessary for successful execution.

Aerial manipulation extends shared-load transport to a cable-suspended payload. The framework in \cite{fink2011planning} combines robot motion planning with configuration constraints and individual control laws to control payload position and orientation. This separates planning for joint payload motion from individual robot control. In this setting, multiple cable forces supply the actuation needed for the desired payload pose. The authors evaluate the approach in simulation and with quadrotor experiments.

System integration can similarly enable collaborative sensing across operating domains. In \cite{lindsay2022collaboration}, an unmanned surface vehicle (USV) hosts a centralized controller and provides the communication connection between an unmanned aerial vehicle (UAV) and an unmanned underwater vehicle (UUV), which cannot communicate directly. The resulting combination of communication, role assignment, and heterogeneous sensing supports a multi-domain sensing capability beyond that of any individual platform.

Heterogeneous coverage combines sensing modalities distributed across a team. The coverage laws in \cite{santos2018coverage, santos2018coveragelimited} associate environmental features with the sensors capable of detecting them and couple robot motion through neighboring robots' states. The latter also uses information about neighbors' sensor modalities to assign sensing responsibilities. For configurations in which no robot carries the full set of required sensors, the team can monitor feature types unavailable to any individual robot. The combined sensing modalities supply the joint capability, while the distributed coverage laws organize their use.

Distributed optimization and MPC provide complementary planning and control tools for these task-level frameworks. The methods discussed in the background section organize information exchange, reference choices, and coupled control decisions. Their relevance to collaboration is in arranging the execution of a task that requires combined capabilities; the required joint capability remains a property of the task and robot team.

\section{Comparisons}\label{sec:comparisons}

\newcolumntype{C}[1]{>{\centering\arraybackslash}p{#1}}
\begin{table*}[!t]
    \centering
    \small
    \renewcommand{\arraystretch}{1.5}
    \caption{Collaborative task settings under Definition~\ref{def:collab}, grouped by centralized, decentralized, and hierarchical decision-making. Within each architectural group, references with the same framework features share a row. A checkmark indicates a feature used in the described setting; a cross indicates its absence from that setting's characterization.}
    
\resizebox{\textwidth}{!}{%
\begin{tabular}{C{2cm}|C{1.5cm} C{1.8cm} C{1.7cm} C{1.4cm} C{0.9cm} C{2.1cm} C{2.1cm}}
        \hline
        \textbf{Papers} & \textbf{Centralized} & \textbf{Decentralized} &
        \textbf{Hierarchical} & \textbf{Ecological} &
        \textbf{HSI} & \textbf{Game Theoretic} & \textbf{Learning-Based} \\
        \hline
\cite{Nguyen23:Mutualistic, nguyen2024scalable, Nguyen25:Mutualisms} & \large{$\checkmark$} & \large{$\times$} & \large{$\times$} & \large{$\checkmark$} & \large{$\times$} & \large{$\times$} & \large{$\times$} \\
\hline
\cite{martin2026, lindsay2022collaboration} & \large{$\checkmark$} & \large{$\times$} & \large{$\times$} & \large{$\times$} & \large{$\times$} & \large{$\times$} & \large{$\times$} \\
\hline
\cite{liu2024coherent} & \large{$\checkmark$} & \large{$\times$} & \large{$\times$} & \large{$\times$} & \large{$\times$} & \large{$\times$} & \large{$\checkmark$} \\
\hline
\cite{santos2018coverage, santos2018coveragelimited} & \large{$\times$} & \large{$\checkmark$} & \large{$\times$} & \large{$\times$} & \large{$\times$} & \large{$\times$} & \large{$\times$} \\
\hline
\cite{notomista2019constraint, lin2025heterogeneous} & \large{$\times$} & \large{$\checkmark$} & \large{$\times$} & \large{$\checkmark$} & \large{$\times$} & \large{$\times$} & \large{$\times$} \\
\hline
\cite{shan2024distributed, liu2025game} & \large{$\times$} & \large{$\checkmark$} & \large{$\times$} & \large{$\times$} & \large{$\times$} & \large{$\checkmark$} & \large{$\times$} \\
\hline
\cite{Zhou2022, Miller22:Stronger, RoCo2024, zhang2020decentralized, dai2024dynamic, dai2025heterogeneous} & \large{$\times$} & \large{$\checkmark$} & \large{$\times$} & \large{$\times$} & \large{$\times$} & \large{$\times$} & \large{$\checkmark$} \\
\hline
\cite{alkilabi2016transport, karam2026altruistic} & \large{$\times$} & \large{$\checkmark$} & \large{$\times$} & \large{$\checkmark$} & \large{$\times$} & \large{$\times$} & \large{$\checkmark$} \\
\hline
\cite{hekmatfar2014cooperative, fink2011planning} & \large{$\times$} & \large{$\times$} & \large{$\checkmark$} & \large{$\times$} & \large{$\times$} & \large{$\times$} & \large{$\times$} \\
\hline
\cite{hussein2022characterization, divband2021designing, karam2026freehand} & \large{$\times$} & \large{$\times$} & \large{$\checkmark$} & \large{$\times$} & \large{$\checkmark$} & \large{$\times$} & \large{$\times$} \\
\hline
\cite{serpiva2021swarmpaint} & \large{$\times$} & \large{$\times$} & \large{$\checkmark$} & \large{$\times$} & \large{$\checkmark$} & \large{$\times$} & \large{$\checkmark$} \\
\hline
\cite{zhang2024coalition} & \large{$\times$} & \large{$\times$} & \large{$\checkmark$} & \large{$\times$} & \large{$\times$} & \large{$\checkmark$} & \large{$\times$} \\
\hline
\end{tabular}
}

    \label{tab:paper_class}
\end{table*}

Table~\ref{tab:paper_class} compares collaborative task settings discussed throughout ``Multi-Robot Collaboration,'' including team allocation, joint execution, and integration of complementary sensing and mobility. Inclusion requires a shared non-adversarial objective, information that couples robot decisions, and a task outcome requiring combined capabilities under Definition~\ref{def:collab}. This includes planning and coalition-formation methods when they arrange capabilities required for an individually unattainable task. The classification concerns the specified task setting, rather than every application of a method. General surveys and enabling tools discussed without such a task setting remain part of the narrative comparison.

Organizational structure is the primary grouping criterion: decisions about joint execution are made by a central planner, by individual robots, or across hierarchical levels. The decentralized column follows this survey's non-centralized convention, including wireless communication and local sensing, and classifies learned policies by execution rather than training. Hierarchical settings separate team planning or supervisory guidance from robot execution. The HSI entries include MICAH's illustrative rescue scenario \cite{hussein2022characterization}, human-guided distributed perception \cite{divband2021designing}, and human-specified geometric outputs. Within each architecture, references with identical framework features share a row.

The framework columns retain the four perspectives discussed in the survey. Ecological control, game-theoretic coalition formation, and learned policies can support different stages of collaboration; evolved collective transport combines ecological inspiration with learning. The HSI entries connect human task specification with complementary aerial and ground capabilities, distributed perception, or collective geometric outputs. SwarmPaint also uses learning to recognize the gestures specifying the task; its learning mark refers to this interface component. Task allocation, heterogeneous coverage, multi-domain system integration, and shared-object manipulation also support collaboration through mechanisms beyond these four perspectives, reflected by unmarked framework columns. The table characterizes the surveyed task settings rather than the relative size or maturity of research areas.

\section{Challenges and Future Research Directions}\label{sec:challenges}

The surveyed mechanisms motivate four research problems: representing task-dependent joint capabilities, preserving guarantees as teams change, selecting collaborative actions with local information, and translating supervisory intent into feasible joint execution. Each problem connects the taxonomy to a specific modeling or design requirement.

\subsection{Models of Task Feasibility and Joint Capability}

How can a model identify which combinations of robot capabilities make a specified task feasible when no individual robot can complete it? The required development is a representation that connects task constraints and individual capabilities to feasible joint actions, while distinguishing team formation from the execution of those actions. The warehouse example illustrates the distinction: selecting a group and realizing its shared transport capability are separate requirements. Such a model should keep the task assumptions fixed when comparing individual and joint feasibility, including any stated reliability requirements.

Game-theoretic and ecological formulations provide starting points for examining how incentives relate to these feasible actions. Equilibrium and efficiency concepts \cite{rosen_ExistenceUniquenessEquilibrium_1965, censor1977pareto} describe strategic outcomes, while altruistic and socially aware approaches examine contributions to collective outcomes at individual cost \cite{Morton09:Altruistic, Toghi22:Social, Valiente22:Robustness, karam2025resource, butler2025hamilton}. A concrete research question is how to connect these objectives to the capabilities required by a task, so that a selected coalition has both an incentive to participate and the means to execute its assigned joint action.

\subsection{Guarantees During Changes in Collaborative Arrangements}

How can stability, safety, and task performance be maintained when robots enter or leave a collaborative arrangement? Existing work on coordinated behavior and resilience provides relevant foundations \cite{olfati2002graph, prorok2021beyond}. The collaborative setting adds dependencies through shared physical interaction, complementary observations, or jointly constrained decisions. Changing membership can therefore change both the applicable control constraints and the capabilities available to complete the task.

A focused problem is to characterize conditions under which a transition between teams preserves safe execution and leaves the remaining task feasible. This requires relating guarantees within an arrangement to the conditions for forming, changing, or ending it. For shared transport, the question concerns the transition between individual motion and joint manipulation; for complementary sensing, it concerns whether the observations needed for the task remain available. The desired result is a set of conditions linking admissible team changes to control guarantees and continued task feasibility.

\subsection{Selection of Partners and Joint Actions with Local Information}

How can a robot decide when to seek a collaborator, which robot to approach, and which joint action to undertake when its observations are incomplete? MARL and CTDE offer tools for learning team behavior \cite{Orr23:Multi_Agent}, but the design problem is to represent these choices explicitly and relate them to the capabilities needed by the current task. A focused question is what information about other robots' capabilities and intentions is sufficient to make compatible participation decisions during decentralized execution.

Networked learning and shared-value methods offer mechanisms for information alignment, as illustrated by consensus-style learning \cite{zhang2018fully} and lane-changing decision-making \cite{Bi_He_Sun_2025}. Learned formation and coverage policies \cite{tolstaya_2020, Bai_2021} and graph-based prediction of goals and interaction intent \cite{goarin2024graph} offer additional representations. The research task is to connect such information and representations to partner selection and jointly feasible actions under partial observability. Evaluation should distinguish successful realization of a required joint capability from improvements in coordination efficiency or learning speed.

\subsection{From Supervisory Intent to Feasible Joint Execution}

How can high-level human guidance be translated into team assignments that respect robot capabilities and the dependencies of a collaborative task? Language-based interaction provides a planning interface \cite{li2024survey}; its integration with collaborative execution calls for a connection between the requested outcome, the capabilities assigned to it, and the information needed to carry it out.

A concrete problem is to determine when a proposed task decomposition can be executed by the available team and when it should be revised or returned for supervisory input. A related question is how authority should shift between the supervisor and robot team as task conditions change. Addressing these questions requires linking planning decisions to execution feedback and specifying intervention conditions in terms of safety, task feasibility, and progress. This makes the trade-off between human oversight and robot autonomy a design problem tied to the collaborative task.

\subsection{Concluding Remarks}

This survey examined cooperation, coordination, and collaboration in multi-robot systems, emphasizing how shared intent, information exchange, and combined capabilities relate to task accomplishment. The reviewed organizational structures and frameworks illustrate different ways of arranging and supporting collaboration. These distinctions may help readers compare approaches while keeping their task requirements and assumptions explicit. The research directions identified here connect this perspective to models of joint capability, guarantees during team changes, decisions under partial observability, and supervisory planning. Progress on these questions may support the design of multi-robot systems that can combine their capabilities reliably as tasks and operating conditions change.

\bibliographystyle{unsrt}
\bibliography{references}

\end{document}